\documentclass{article}
\usepackage{graphicx}
\usepackage{amsmath}
\usepackage{hyperref}
\usepackage{authblk}
\begin{document}
\title{Two-pion source analysis in small to large systems at SPS energies}

\author[1,2]{Barnabás Pórfy}
\author[3]{Zsombor Marcell Rácz}
\affil[1]{Department of Atomic Physics, ELTE E\"otv\"os Lor\'and University, Budapest, Hungary}
\affil[2]{HUN-REN Wigner Research Centre for Physics, Budapest, Hungary}
\affil[3]{Kossuth Lajos High School, Budapest, Hungary}
\maketitle

\begin{abstract}
Over the past decade, femtoscopic measurements using Lévy-stable source functions have been carried out in collision systems ranging from small to large at SPS, RHIC, and LHC energies. In recent years, these have been accompanied by a number of phenomenological studies of the same systems and energies. We contribute to this line of work by investigating the shape of two-pion pair-source distributions in central $^7$Be+$^9$Be and $^{129}$Xe+$^{139}$La collisions simulated with the UrQMD model in cascade mode. To make the comparison with measurements meaningful, kinematic selections similar to the experimental ones are applied to the simulated events and particles. One-dimensional Lévy fits are performed, and the resulting source parameters are studied as functions of pair transverse mass, collision energy, and system size. We find that the simulated sources are well described by Lévy distributions in all systems considered, which offers a common phenomenological picture of the space-time structure of pion emission. Finally, we compare our findings with results from an intermediate system, Ar+Sc. The results serve as a purely hadronic reference for past and upcoming femtoscopic measurements of the NA61/SHINE experiment.
\end{abstract}


\section{Introduction}
High energy heavy-ion collisions produce the quark--gluon plasma (QGP), and characterizing its properties is a central aim of many experimental programs. The NA61/SHINE experiment at the CERN SPS addresses this question by scanning a two-dimensional parameter space spanned by collision energy and system size~\cite{Abgrall:2014fa}. Femtoscopy, the study of quantum-statistical correlations between identical particles~\cite{Goldhaber:1959mj, Goldhaber:1960sf}, is well suited for such a scan, since it gives access to the femtometer-scale space-time extent of particle emission~\cite{Lisa:2005dd, Csorgo:1999sj}.

For a long time, the shape of the particle-emitting source was assumed to be Gaussian~\cite{Csorgo:1994fg, Akkelin:1995gh}. Measurements performed in various systems and at various energies~\cite{PHENIX:2017ino,NA61SHINE:2023qzr,CMS:2023xyd,Kincses:2024sin,PHENIX:2024vjp} point towards the more general, Lévy-stable shape~\cite{Csorgo:2003uv,Csanad:2007fr} featuring a power-law tail. There exist several mechanisms that are known to generate such power-law tails, for example, jet fragmentation~\cite{Csorgo:2004sr} and critical behavior~\cite{Csorgo:2005it}. Anomalous diffusion in the hadronic stage, referred to as a Lévy walk~\cite{Kincses:2024lnv}, is considered the most prominent of these.

Simulations are essential to complement any experimental measurements. In the case of femtoscopy, this is because an experiment can only access the pair source indirectly, from momentum correlations, whereas an event generator can record the space-time coordinates of the last interaction of every final-state particle, thus, the pair source can be reconstructed directly. Building on this, the present work studies like-sign pion pair sources in $^7$Be+$^9$Be and $^{129}$Xe+$^{139}$La collisions, representing small and large systems, respectively. The events are simulated with UrQMD~\cite{Bass:1998ca, Bleicher:1999xi} at beam momenta of 13$A$--150$A$ GeV/$c$ ($\sqrt{s_\text{NN}} \approx 5$--$17$ GeV), matching the NA61/SHINE strong interactions program. The angle-averaged sources are fitted with one-dimensional Lévy distributions~\cite{LevyCode}, following the methodology of earlier phenomenological studies~\cite{Kincses:2024lnv, Kincses:2025iaf, Kincses:2025izu, Huang:2025edi, Porfy:2025sev, Molnar:2026vgi, Kincses:2026tfa}, and the parameters are examined as functions of the transverse mass ($m_\text{T} = \sqrt{K_\text{T}^2+m^2}$), the collision energy, and the system size.

The paper is organized as follows. Section~\ref{s:formalism} briefly summarizes femtoscopy theory, Section~\ref{s:analysis} describes the simulations and the source reconstruction, Section~\ref{s:results} presents the results, and Section~\ref{s:summary} gives a summary and outlook.

\section{Femtoscopy}\label{s:formalism}
Let $S(x,p)$ denote the single-particle source function, i.e., the probability density of emitting a particle with four-momentum $p$ from the space-time point $x$. Two-particle measurements are sensitive to the pair-source distribution (or spatial pair distribution), which is the autoconvolution of $S$:
\begin{equation}\label{e:pairsource}
	D(r,K) = \int d^4\sigma \, S(\sigma + r/2,K)\,S(\sigma - r/2,K),
\end{equation}
where $r = x_1 - x_2$ is the space-time distance of the pair, $\sigma = (x_1 + x_2)/2$ is the pair center of mass, and $K = (p_1 + p_2)/2$ is the average pair momentum. The pair source is linked to the measurable two-particle correlation function via
\begin{equation}\label{e:corrfunc}
	C_2(q,K) = \int d^4r \, D(r,K)\,|\Psi^{(2)}_q(r)|^2,
\end{equation}
where $\Psi^{(2)}_q$ is the pair wave function and $q = p_1 - p_2$ is the relative momentum. Both Eqs.~\eqref{e:pairsource} and~\eqref{e:corrfunc} depend on $K$, however, this dependence is expected to be weaker than the dependence of $C_2$ on $q$~\cite{Lisa:2005dd}, and in practice it is taken into account by measuring the source parameters as a function of $K$.

In three dimensions, the Lévy-stable distribution is defined as:
\begin{equation}
	\mathcal{L}(\alpha, R^2; \vec{r}) = \frac{1}{(2\pi)^3}
\int d^3\vec{\zeta} \, e^{i \vec{\zeta} \cdot \vec{r}} \, e^{-\frac{1}{2} \left| \vec{\zeta}^T R^2 \vec{\zeta} \right|^{\alpha/2}},
\label{e:Levydef}
\end{equation}
and the pair source is parametrized as $D(\vec{r}) = \mathcal{L}(\alpha, R^2; \vec{r})$. In the above expression, $\vec{\zeta}$ is the integration variable, $\alpha$ is the Lévy stability index, and $R^2$ is the matrix of scale parameters. In the Bertsch--Pratt frame~\cite{Bertsch:1988db, Pratt:1990zq}, $R^2$ is usually taken to be diagonal, $R^2 = \text{diag}\left(R_{\text{out}}^2, R_{\text{side}}^2, R_{\text{long}}^2\right)$. The case $\alpha = 2$ corresponds to a Gaussian and $\alpha = 1$ to a Cauchy distribution, while any $\alpha < 2$ leads to a power-law. A useful property of the stable distributions is that convolving two Lévy-shaped single-particle sources yields a pair source of the same shape, with the same stability index~\cite{PHENIX:2017ino}. Furthermore, the one-dimensional version of Eq.~\eqref{e:Levydef} is as follows: 
\begin{equation}\label{e:1DLevy}
    \mathcal{L}^{1D}(\alpha, R; \vec{r}) = \frac{1}{(2\pi)^3}
\int d^3\vec{\zeta} \, e^{i \vec{\zeta} \cdot \vec{r}} \, e^{-\frac{1}{2} \left| \vec{\zeta} R\right|^{\alpha}},
\end{equation}
which is then extended with $\lambda$ as $\lambda\mathcal{L}^{1D}({\alpha, R;\vec{r}})$, where $\lambda$ is the intercept parameter, which is connected to the core-halo picture~\cite{Bolz:1992hc,Csorgo:1994in}.

\section{Simulation and analysis}\label{s:analysis}

\subsection{Event generation}
Events were generated with UrQMD 3.4 in cascade mode, i.e., without a hydrodynamic stage. As in Ref.~\cite{Porfy:2025sev}, the code was modified, based on Ref.~\cite{Uzhinsky:2013ata}, so that $\eta$ mesons are allowed to decay but it is important to highlight that UrQMD considers kaons, $\Sigma$, $\Xi$, and $\Omega$ baryons as stable. The simulated systems, beam momenta, and event statistics are listed in Table~\ref{tab:generatordet}. The 0--10\% centrality class is defined via the impact parameter derived from minimum bias simulations.

\begin{table}[h]
 \centering
 \caption{Overview of the simulated data sets.}
 \label{tab:generatordet}
    \begin{tabular}{l|c|c}
    \hline
    System & $p_\text{beam}$ [$A$ GeV/$c$] & $N_\text{events}$ \\ \hline\hline
    $^7$Be+$^9$Be & 150 & 100\,000 \\ \hline
    $^{129}$Xe+$^{139}$La & 13, 19, 30, 40, 75, 150 & \begin{tabular}[c]{@{}c@{}}10\,000\\ (100\,000 at 13$A$ and 19$A$)\end{tabular}
    \end{tabular}
\end{table}

\subsection{Reconstruction of the pair source}
The spatial pair distributions are studied in the longitudinally comoving system (LCMS), in which the longitudinal component of the pair momentum vanishes, and its components are ``out'', ``side'', and ``long'' directions~\cite{Bertsch:1988db, Pratt:1986ev}. Following Ref.~\cite{Kincses:2024lnv}, we can write the pair separation four-vector as $(t, \vec{r})$, and taking, $\vec{\rho} = \vec{r} - (\vec{K}/K_0)\, t$, we can arrive to the LCMS components as:
\begin{align}
	\rho_\text{out}^\text{LCMS} &= r_x \cos \varphi + r_y \sin \varphi - \frac{K_\text{T}}{K^2_0 - K_z^2} \left(K_0t - K_z r_z\right), \\
	\rho_\text{side}^\text{LCMS} &= -r_x \sin \varphi + r_y \cos \varphi, \\
	\rho_\text{long}^\text{LCMS} &= \frac{K_0 r_z - K_z t}{\sqrt{K^2_0 - K_z^2}},
\end{align}
with $K_\text{T}^2 = K_x^2 + K_y^2$, $\cos \varphi = K_x/K_\text{T}$, and $\sin \varphi = K_y/K_\text{T}$. Near mid-rapidity, the off-diagonal elements of $R^2$ in Eq.~\eqref{e:Levydef} are expected to vanish~\cite{Chapman:1994yv, PHENIX:2004yan}, and the source becomes close to spherically symmetric ($R_{\text{out}} \approx R_{\text{side}} \approx R_{\text{long}}$)~\cite{PHENIX:2017ino}.

In this paper, the analysis is restricted to the angular average $\rho = |\vec{\rho}\,|$, and the resulting $D(\rho)$ is fitted with the one-dimensional Eq.~\eqref{e:1DLevy}:

A single simulated event does not contain enough pairs for a stable fit~\cite{Korodi:2022ohn,Kincses:2022eqq} in either system. Therefore, pairs are accumulated over blocks of events, each block is fitted separately, and the parameters are averaged over the blocks, as in Refs.~\cite{Kincses:2025izu, Huang:2025edi, Porfy:2025sev, Molnar:2026vgi, Kincses:2026tfa}. To make the simulation results comparable to the experimental measurements, we apply event and track selections similar to the experimental analyses~\cite{NA61SHINE:2023qzr, Porfy:2024jsu}. The block sizes, kinematic and relative-momentum cuts, and $K_\text{T}$ binning are detailed in Table~\ref{tab:kinematic_cuts}.

\begin{table}[h]
 \centering
 \caption{Event-block sizes, single-particle and pair selections, and $K_\text{T}$ binning used for each system.}
 \label{tab:kinematic_cuts}
 \resizebox{\textwidth}{!}{%
 \begin{tabular}{l|c|c|c|c|c}
 \hline
 System & $N_\text{event}$ block sizes & Rapidity& $p_\text{T}$ [GeV/$c$] & $q_\text{LCMS}$ [GeV/$c$] & $K_\text{T}$ [GeV/$c$] \\ \hline\hline
 $^{7}$Be+$^{9}$Be & 5000 & $|y| < 1$ & $<1$ & $< \sqrt{m_\text{T} \cdot 150 \text{ MeV}/c^2}$ & 7 bins in $[0, 0.7]$ \\ \hline
 $^{129}$Xe+$^{139}$La & \begin{tabular}[c]{@{}c@{}}5000 ($p_\text{beam} \ge 30A$)\\ 50\,000 ($p_\text{beam} < 30A$)\end{tabular} & $|y| < 1$ & $<1$ & $< \sqrt{m_\text{T} \cdot 150 \text{ MeV}/c^2}$ & 10 bins in $[0, 1]$ \\ 
 \end{tabular}}
\end{table}

\subsection{Systematic uncertainties}
The statistical uncertainties of the event-block averaged parameters are negligible; the uncertainty is instead dominated by systematic effects. We considered the relative-momentum selection as a source of uncertainty. Beside the default cut shown in Table~\ref{tab:kinematic_cuts}, we studied a strict variation, $q_\text{LCMS} < \sqrt{m_\text{T} \cdot 100 \text{ MeV}/c^2}$, and a loose variation, $q_\text{LCMS} < \sqrt{m_\text{T} \cdot 200 \text{ MeV}/c^2}$. For a parameter $P$, the difference of the varied values $P^{\rm var}_{i,j}$ from the default value $P^{\rm def}$ are combined separately in the upward and downward directions ($j = +,-$):
\begin{align}
 \sigma^2_{i,j}[P] &= (P^\text{def} - P^{\rm var}_{i,j})^2, \\
 \sigma^\text{tot}_j[P] &= \sqrt{\sum^n_{i=1} \sigma^2_{i,j}[P]}.
\end{align}
The resulting asymmetric uncertainties, $\sigma^\text{tot}_\pm$, are shown as bands in the figures.

\section{Results}\label{s:results}
All results refer to 0--10\% central collisions simulated with UrQMD 3.4 at the NA61/SHINE beam momenta. The parameters are presented as functions of the pair transverse mass and of $\sqrt{s_\text{NN}}$. Note that for the collision energy dependence, we added in addition the results (Ar+Sc) from Ref.~\cite{Porfy:2025sev}. It is also important to highlight that the Be+Be and Xe+La sources were analyzed in one dimension, whereas the Ar+Sc source was analyzed in three dimensions.

\subsection{Lévy stability index}
\begin{figure}[h]
\centerline{\includegraphics[width=1.0\textwidth]{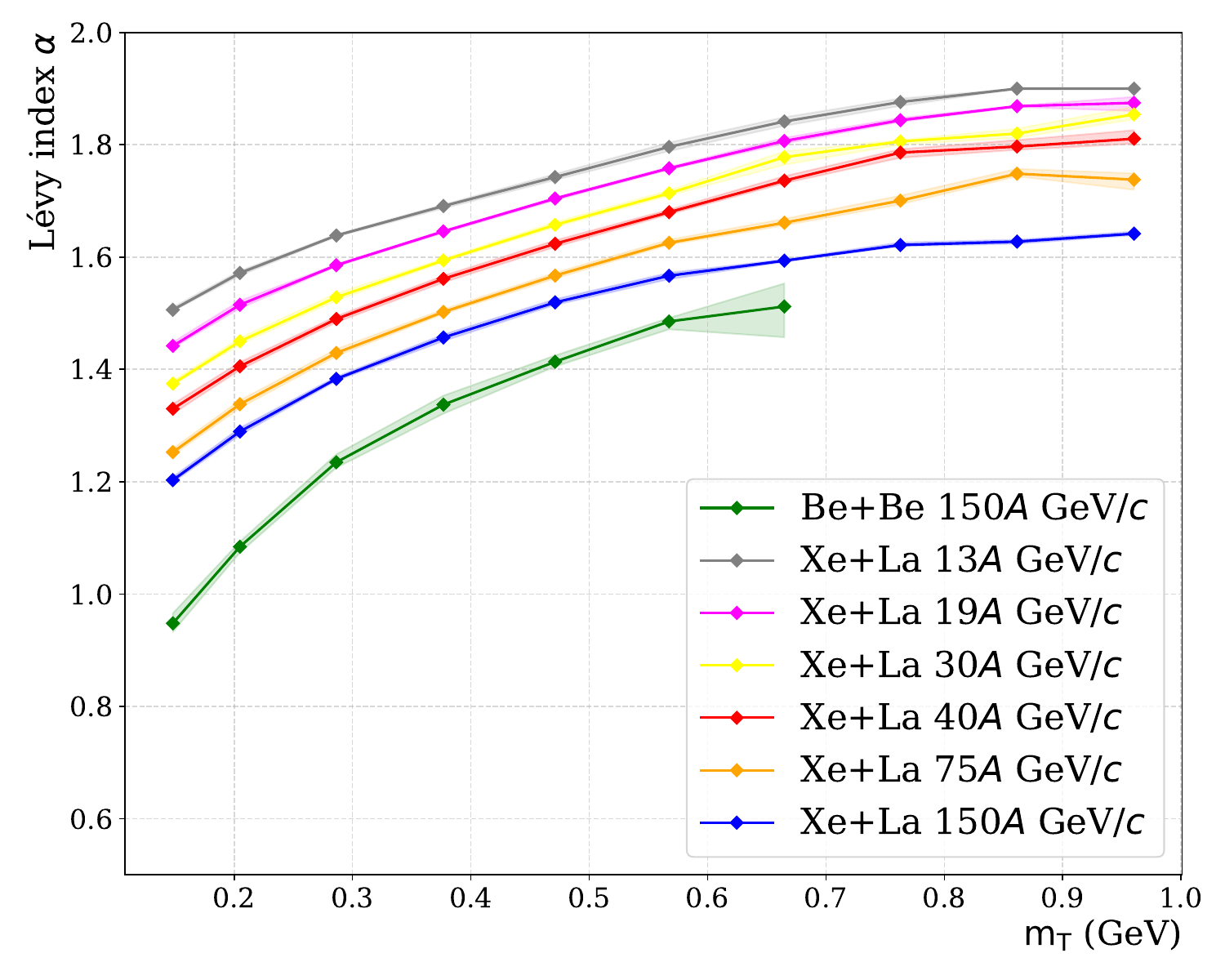}}
\caption{Lévy stability index $\alpha$ versus $m_\text{T}$ in 0--10\% central Be+Be and Xe+La collisions. Bands show the systematic uncertainties.}
\label{fig:XeLaBeBe_A}
\end{figure}

The $m_\text{T}$ dependence of $\alpha$ in Be+Be and Xe+La is visible in Fig.~\ref{fig:XeLaBeBe_A}. In every case, $\alpha$ increases with $m_\text{T}$, and when comparing different collision energies, the value systematically decreases as the collision energy rises. A possible explanation is a stronger Lévy-walk phenomenon, characteristic of the hadronic rescattering at higher energies~\cite{Kincses:2024lnv}. More abundant production of heavier resonances, those that decay in UrQMD, may enhance the power-law tail as well. 

\begin{figure}[h]
\centerline{\includegraphics[width=1.0\textwidth]{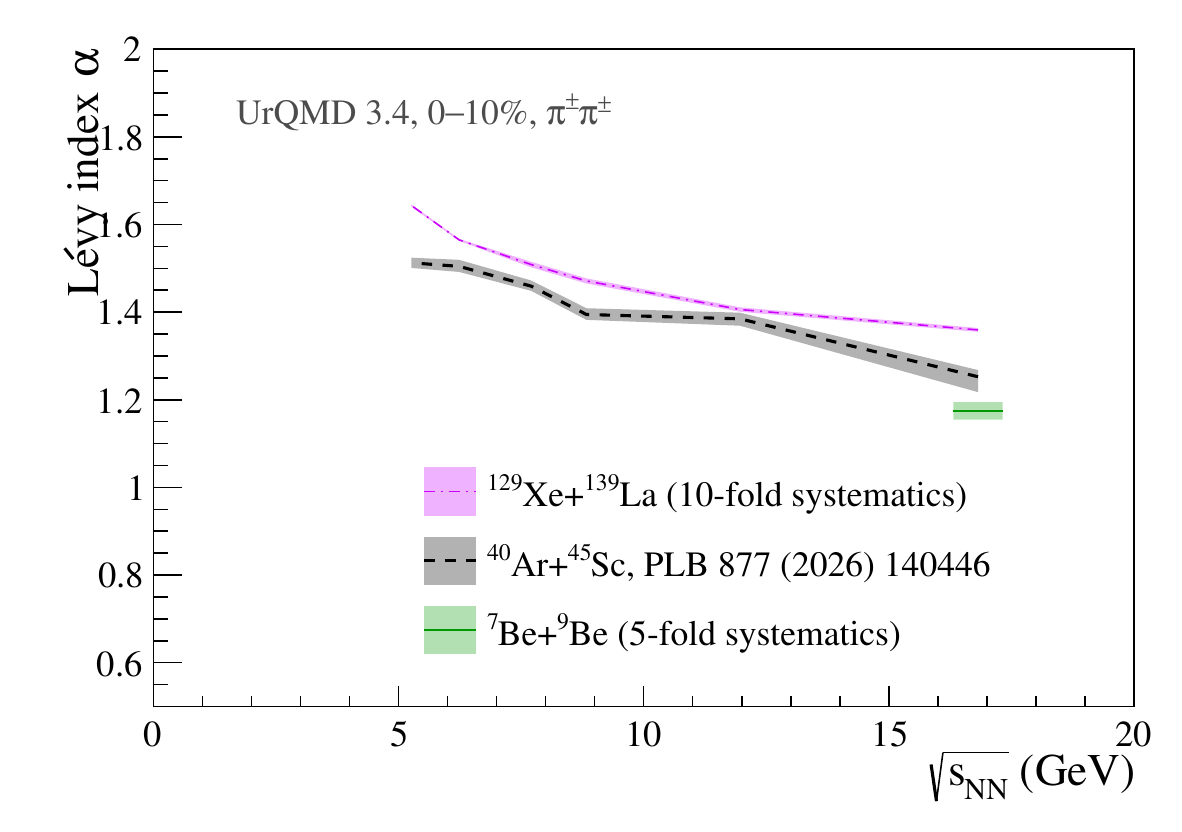}}
\caption{Lévy stability index $\alpha$ versus $\sqrt{s_\text{NN}}$ in 0--10\% central Be+Be, Ar+Sc, and Xe+La collisions, averaged over all $K_\text{T}$. Bands show the systematic uncertainties. The single Be+Be point is drawn with an artificial horizontal width for visibility.}
\label{fig:XeLaBeBe_sNN_A}
\end{figure}

The energy dependence of $\alpha$ is summarized in Fig.~\ref{fig:XeLaBeBe_sNN_A}. The Ar+Sc and Xe+La values separate at the lowest and highest energies and approach each other in between. At the top SPS energy, the values are ordered by system size, with the smallest $\alpha$ in Be+Be, followed by Ar+Sc, and the largest in Xe+La. The simulated results differ from the NA61/SHINE measurement in Be+Be, which yields $\alpha \approx 1$~\cite{NA61SHINE:2023qzr}.


\subsection{Lévy scale parameter}
\begin{figure}[h]
\centerline{\includegraphics[width=1.0\textwidth]{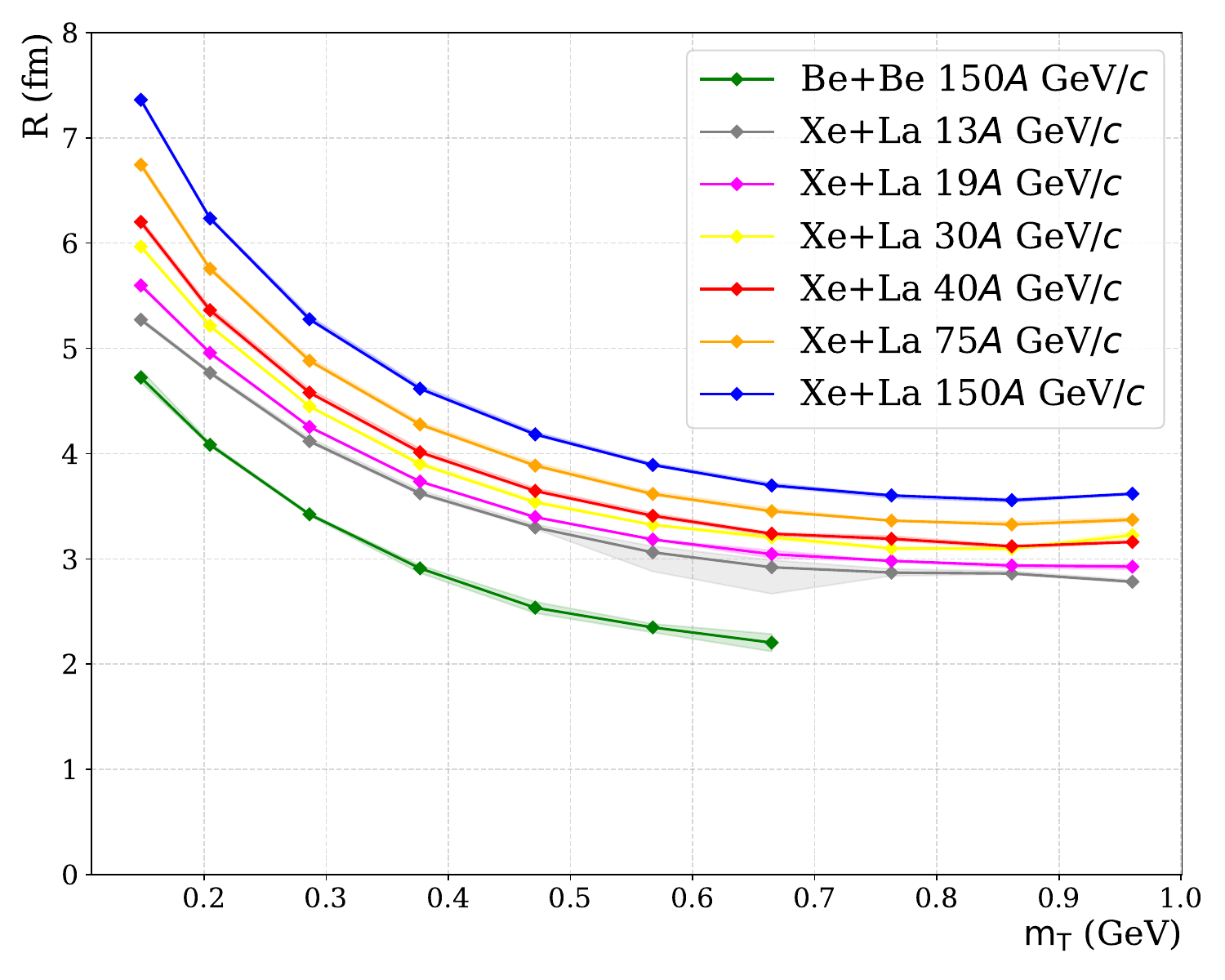}}
\caption{Lévy scale parameter $R$ versus $m_\text{T}$ in 0--10\% central Be+Be and Xe+La collisions. Bands show the systematic uncertainties.}
\label{fig:XeLaBeBe_R}
\end{figure}

The scale parameter $R$, which is related to the length of homogeneity, is shown in Fig.~\ref{fig:XeLaBeBe_R}. It falls steeply with $m_\text{T}$, usually understood as a signature of an expanding source and collective flow~\cite{PHENIX:2004yan}. This was described in hydrodynamic models~\cite{Csorgo:1995bi}, and interestingly, the results in this work indicate that it is also present, at least to some degree, in the hadronic cascade of UrQMD. The value of $R$ increases with collision energy. The Be+Be source is considerably smaller than the Xe+La source, which can be attributed to differences in the lifetime and the radial expansion velocity of the two systems. This difference can also be attributed to different initial geometric size as well.

\begin{figure}[h]
\centerline{\includegraphics[width=1.0\textwidth]{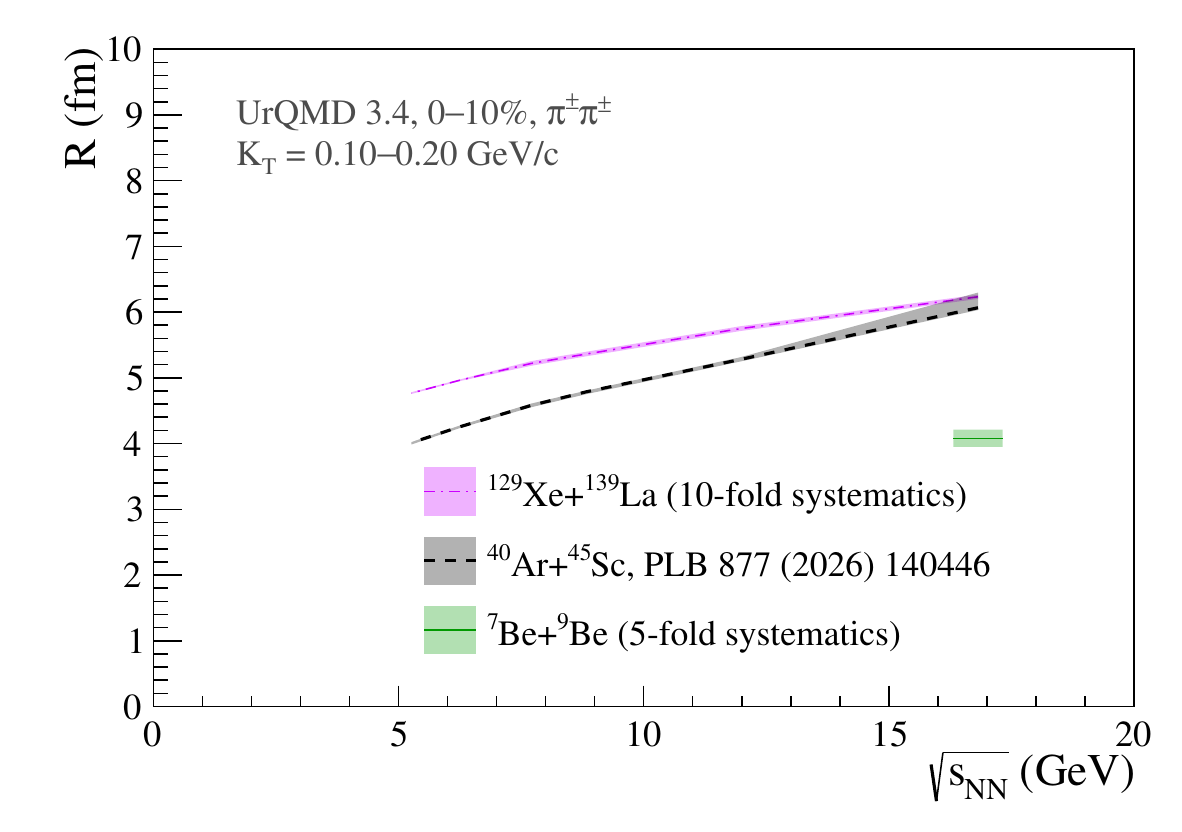}}
\caption{Lévy scale parameter $R$ versus $\sqrt{s_\text{NN}}$ for $K_\text{T} = 0.10$--$0.20$ GeV/$c$ in 0--10\% central Be+Be, Ar+Sc, and Xe+La collisions. Bands show the systematic uncertainties. The single Be+Be point is drawn with an artificial horizontal width for visibility. The systematic uncertainties for Xe+La were artifically increased 10 times, and 5 times for Be+Be, to enhance visibility. For Ar+Sc, the direction-averaged $\bar{R} = \sqrt{(R^2_\text{out} + R^2_\text{side} + R^2_\text{long}) / 3}$ is shown.}
\label{fig:XeLaBeBe_sNN_R}
\end{figure}

Figure~\ref{fig:XeLaBeBe_sNN_R} shows the energy dependence of $R$ for $K_\text{T} = 0.10$--$0.20$ GeV/$c$. At low energies, a clear system size ordering is visible with Xe+La radii above Ar+Sc. At the highest energy, the two systems coincide, while the Be+Be result is smaller than both systems.


\subsection{Correlation strength}
\begin{figure}[h]
\centerline{\includegraphics[width=1.0\textwidth]{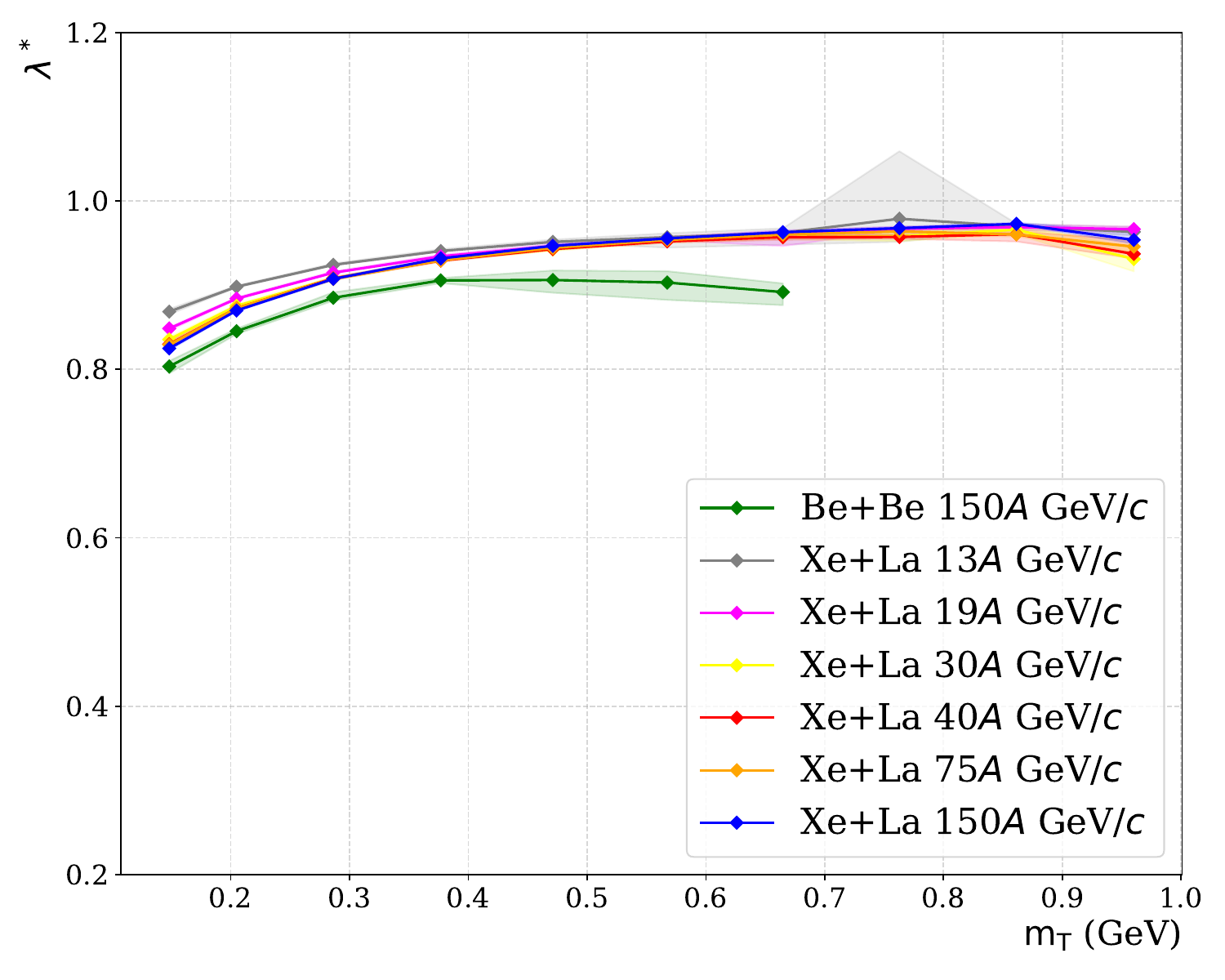}}
\caption{Correlation strength $\lambda^*$ versus $m_\text{T}$ in 0--10\% central Be+Be and Xe+La collisions. Bands show the systematic uncertainties.}
\label{fig:XeLaBeBe_L}
\end{figure}

The correlation strength is presented in Fig.~\ref{fig:XeLaBeBe_L}. A slight increase is apparent at low $m_\text{T}$, which then flattens at higher $m_\text{T}$, with almost no collision energy dependence. Pions from certain weak decays would populate the halo and thus reduce the parameter value. The absence of these decays leads to an overestimated value, denoted as $\lambda^*$ to distinguish it from the experimentally measured $\lambda$. An estimate based on the branching ratios of the above mentioned decays into pions suggests that the physical $\lambda$ would be approximately 92--93\% of the values shown.

\section{Summary and outlook}\label{s:summary}
We have analyzed two-pion pair-source distributions in central Be+Be and Xe+La collisions at SPS energies ($\sqrt{s_\text{NN}} \approx 5$--$17$ GeV), simulated with the UrQMD model in cascade mode, in order to provide a hadronic reference for femtoscopy measurements at NA61/SHINE. In all systems, the simulated sources are compatible with a Lévy shape. 

Both $\alpha$ and $R$ show a clear system-size dependence, whereas $\lambda^*$ is seemingly constant with respect to the system size. With increasing $m_\text{T}$, $R$ decreases, while $\alpha$ and $\lambda^*$ increase. As a function of collision energy, Ar+Sc and Xe+La show similar trends, while at top SPS energy, Be+Be is clearly distinct from both.

Finally, the next step is extending the model with a hydrodynamic stage, and the present results provide the baseline against which such extensions and future measurements can be compared.

\section*{Acknowledgments}
This work was supported by the NKFIH grant NKKP ADVANCED 152097. Furthermore, Barnabás Pórfy was supported by the DKÖP-23 Doctoral Excellence Program of the Ministry for Culture and Innovation and by the EKÖP-26 University Research Scholarship Program of the Ministry of Science and Technology from the source of the National Research, Development and Innovation Fund. Zsombor Marcell Rácz was supported by the ELTE Talentum Program.

\bibliographystyle{ws-ijmpa.bst}
\bibliography{wpcf}
\end{document}